\documentclass[aps,pra,twocolumn,twocolumn]{revtex4-1} 
\usepackage{amsmath}
\usepackage{latexsym}
\usepackage{graphicx}
\usepackage{amsfonts}
\usepackage{braket}
\usepackage{natbib}
\usepackage{amssymb}
\usepackage{fancyhdr}
\usepackage[utf8]{inputenc}
\usepackage{dsfont}
\usepackage{colortbl}
\usepackage{xcolor}
\usepackage{subfigure}
\usepackage{psfrag}
\usepackage[colorlinks=true,citecolor=blue,linkcolor=blue]{hyperref}
\usepackage{tikz}
\usepackage{ulem}
\usepackage{bbold}
\usepackage{bbm}
\newcommand\bbone{\ensuremath{\mathbbm{1}}}
\usetikzlibrary{arrows,shapes}

\newcommand{\La}{$\Lambda$ }
\newcommand{\ketbra}[2]{\ket{#1}\hspace*{-0.12cm}\bra{#2}}

\begin{document}

\title{A complete Randomized Benchmarking Protocol accounting for Leakage Errors}
\author{T. Chasseur}
\affiliation{Theoretical Physics, Universit\"at des Saarlandes, 66123 Saarbr\"ucken, Germany}
\author{F.K. Wilhelm}
\affiliation{Theoretical Physics, Universit\"at des Saarlandes, 66123 Saarbr\"ucken, Germany}

\begin{abstract}
Randomized Benchmarking allows to efficiently and scalably characterize the average error of an unitary 2-design such as the Clifford group $\mathcal{C}$ on a physical candidate for quantum computation, as long as there are no non-computational leakage levels in the system. We investigate the effect of leakage errors on Randomized Benchmarking induced from an additional level per physical qubit and provide a modified protocol that allows to derive reliable estimates for the error per gate in their presence. We assess the variance of the sequence fidelity corresponding to the number of random sequences needed for valid fidelity estimation. Our protocol allows for gate dependent error channels without being restricted to perturbations. 
We show that our protocol is compatible with Interleaved Randomized Benchmarking and expand to benchmarking of arbitrary gates. This setting is relevant for superconducting transmon qubits, among other systems. 
\end{abstract}

\maketitle

\section{Indroduction}
In the wake of recent advances in experimental implementations of quantum gates on physical qubits \cite{Barends2014,Chow2014,Harty2014}, characterizing the fidelity of those gates efficiently and accurately becomes increasingly important. The original standard approach to achieving this is quantum process tomography (QPT) \cite{Nielsen2000} which is not reliable as it cannot discriminate gate errors from those of state preparation and measurement (SPAM) \cite{Knill2008}. Neither is it practical due to the number of measurements needed, which is exponential in the number of subsystems. The Randomized Benchmarking (RB) protocol is a scalable and SPAM independent method to benchmark unitary 2-designs such as Clifford gates by characterizing the fidelity of sequences of random gates \cite{Knill2008,Magesan2011}. Interleaved Randomized Benchmarking (IRB) provides a means to estimate the average fidelity of a single gate of the group up to a uncertainty defined by the group fidelity \cite{Magesan2012}. As the RB protocol is used  not only for benchmarking but for closing the loop in experiment optimal control \cite{Egger2014,Kelly2014} it becomes increasingly imperative for all possible errors connected to this protocol.

Many promising candidates for implementing qubits on which RB is applied such as superconducting qubits \cite{Barends2014,Gambetta2012}, NV-centers \cite{Dolde2014}, trapped ions \cite{Knill2008} or neutral atoms \cite{Xia2015} are not natural two level systems making leakage into additional levels a viable error source. Since the physical qubit is protected from unwanted interactions with the environment so are in many cases the leakage levels. Therefore leakage is not accounted for by standard RB, not least because it is a non-Markovian process. 
Furthermore standard RB treats gate dependencies only in the regime of small deviations from a predominant gate independent error. We investigate leakage arising from an additional level per qubit as first discussed by Epstein \emph{et al.} \cite{Epstein2014} in the single qubit case. We provide a generalization of their RB protocol accounting for leakage errors as well as strongly gate dependent error channels. We investigate the number of random sequences needed for a 
good estimate of the fidelity and show that IRB still works--even for gates not in the unitary 2-design.

This work is structured in the following way. We introduce the  protocol in section \ref{sec:prot} and derive the associated fidelity model in section \ref{sec:fidel}. There, we  first make an additional assumption on the Clifford gates and discuss in \ref{subsec:unlike} how useful information can be extracted even if these are not satisfied. In sec. \ref{subsec:spam} we discuss the influence of SPAM errors. Gate dependent errors are treated in section \ref{sec:gdep} and section \ref{sec:irb} covers IRB.

\section{Modified Protocol}
\label{sec:prot}
The RB protocol allows to efficiently estimate  the average fidelity of a unitary 2-design like the N-qubit Clifford group $\mathcal{C}_N$, i.e., find the average fidelity of all Clifford gates. For that one applies a sequence of $y$ random gates and then inverts them with a final gate. The final gate can  be easily calculated on a classical computer according to the Gottesmann-Knill theorem \cite{Gottesman1999}.  It can be shown that the resulting quantum channel approaches the $y$-fold application of a depolarizing error channels \cite{Knill2008,Magesan2011,Dankert2009} and can be fitted to an exponential decay in $y$. A major consequence of the application of RB sequences is that there is no coherent interference between errors meaning here that the error channel does not maintain any well-defined phase relations between different states. This is not the case if one considers transitions to a third, non-computational level in any physical qubit  while still only applying {\em qubit} Clifford gates. For a single qubit, Epstein \emph{et al.} proposed the simple procedure of randomly inserting phase factors $\pm1$ on the third level to reach the required phase randomization \cite{Epstein2014}. The multi-qubit generalization is to randomly insert these phase factors on every qubit independently, thus destroying any phase relation. Including these operations extends the Clifford group $\mathcal{C}$ into the altered Clifford set $\mathcal{C}^\ast$ 
\begin{align}
{\mathcal C}_N^\ast=\mathcal{C}_N \times\left\{\bbone_2\oplus(\pm1) \right\}^{\otimes N}\label{eqn:cstar}
\end{align}
whose size is increased by a factor of $2^N$ for N physical qubits used in our RB protocol, and $\mathcal{C}_N$ is the Clifford group for the corresponding $N$ qubits. It is not necessarily a group since the effect of every implemented gate on the non-computational subspace can be arbitrary. Therefore $\mathcal{C}^\ast$ does not have to be closed or contain an inverse  or even neutral element. As will be derived in section \ref{sec:fidel} the parameterized fidelity one needs to fit is
\begin{align}
\Phi_y=\sum_i a_i \lambda_i^y \label{eqn:model}
\end{align}
with $\lambda_i$ eigenvalues of an operator related to the error and the $a_i$ define the measurement functional in its eigenbasis, they contain the SPAM errors as one contribution. At least a single $\lambda_i$ is unity and the number of different $\lambda_i$ is smaller than $3^N$. The obtained average fidelity then sums up to $\sum_i a_i \lambda_i/\sum_i a_i$ as the average error of applying a single  gate.

\section{Deriving the Fidelity Model}
\label{sec:fidel}
In this section, we derive a model for the error per gate, neglecting state preparation and measurement errors as well as the error of the inverting gate for simplicity. The effect of this approximation will be discussed  in \ref{subsec:spam}. 
Adding a phase factor on the leakage level as required in the previous section is practically straightforward, for example by waiting the appropriate an amount of time. This time is set through the qutrit anharmonicity $\delta\omega$, i.e., the difference in Bohr frequency between the leakage transition and the working transition, as $\tau=\phi/\delta\omega$ to achieve a phase $\phi$. This extends easily to multiple qubits if these are frequency-tunable \cite{Mariantoni2011}. Using this extended Clifford set $\mathcal{C}^\ast$ of equation (\ref{eqn:cstar}) removes all phase relation between different energy levels, thus, following the same reasoning as in other discussions of RB such as Ref. \cite{Epstein2014} all resulting density matrices are diagonal, given by
\begin{align}
\mathcal{H}_d=\left\{\begin{pmatrix} 1 & 0 & 0 \\ 0 & 0 & 0 \\ 0 & 0 & 0 \end{pmatrix},\begin{pmatrix} 0 & 0 & 0 \\ 0 & 1 & 0 \\ 0 & 0 & 0 \end{pmatrix},\begin{pmatrix} 0 & 0 & 0 \\ 0 & 0 & 0 \\ 0 & 0 & 1 \end{pmatrix}\right\}^{\otimes N}.
\end{align}
Using the Frobenius inner product $\braket{\rho_1|\rho_2}={\rm Tr}[\rho_1^\dagger\rho_2]$ and representing diagonal density matrices as vectors and superoperators as matrices the measured fidelity of the RB protocol for gate independent error channel \La is

\begin{align}
\Phi_y&= \frac{1}{{\sharp \mathcal{C}^*}^y}\sum_{\{C_j\}\in{\mathcal{C}^*}^y} \bra{\rho_0} C_{y+1}\prod_{j=y}^1 (\Lambda C_j) \ket{\rho_0}\label{eqn:fid0}\\
&= \frac{1}{{\sharp \mathcal{C}^*}^y}\sum_{\{C_j\}\in{\mathcal{C}^*}^{y}} \bra{\rho_0}  C_{y+1}\prod_{j=y}^3 (\Lambda C_j)~ C_2 C_1\times\notag\\
 &\hspace{1cm}\times C_1^{-1} C_2^{-1} \Lambda C_2 C_1~ C_1^{-1} \Lambda C_1 \ket{\rho_0}\text{.}
\end{align}

$\sharp$ denotes the cardinality and neither the elements of $\mathcal{C}^*$ nor error channels are unitary in this representation. The product indices are counted down in order to ensure that the lower index random Cliffords $C_j$ are applied first. For the standard RB restricted to the computational subspace the group properties of $\mathcal{C}$ yield that

\begin{align}
\sum_{C_2} f(C_2 C_1)=\sum_{C_2} f(C_2)\label{eqn:prop}
\end{align}
for any $C_1 \in \mathcal{C}$ and any function $f$.  Including a third level any unitary operation is no longer fully determined  by its effect on the computational subspace but by the $3^N-2^N$ dimensional non computational subspace as well. Since any gate is engineered with respect to how it acts on the first the effect on the  latter may be arbitrary as long as the subspaces are kept disconnected. Thus, in general, the set $\mathcal{C}^\ast$  is not be a group since it is neither closed nor does it contain inverse or neutral elements {\em a priori}. As equation (\ref{eqn:prop}) relies on the group property it is therefore no longer valid for general $\mathcal{C}^*$ rendering a twirl over \La no longer possible. We will solve this problem in section  \ref{subsec:unlike}.

A gate on a qutrit is considered to be a perfect single qubit gate if it has perfect gate fidelity in the qubit subspace and leaves a phase shift on the leakage level \cite{Rebentrost2009}. Since the random phase shifts randomize that phase we can effectively treat every single qubit gate as if it would act as the identity on the leakage level and the single qubit Clifford set $\mathcal{C}_1^\ast$ as if were a group  in the sense that after the randomization, group properties are recovered within the RB protocol.
 
This argument does not hold in the same form for  more than a single qubit. On the other hand, the multi-qubit Clifford set $\mathcal{C}_N^*$ is generated by its single qubit generators plus at least one entangling Clifford gate between arbitrary qubits. If one presumes that the entangling gate induces a mere phase change on the leakage levels as well -- which is equivalent to it being diagonal in the non computational subspace -- then this means the whole Clifford set has this same property. As above, since phases are averaged to zero by their randomization the set of Clifford gates can be treated as a group as for further calculations. In the following we call such a set a twirl design and it yields a fidelity of
\begin{align}
\Phi_y&= \frac{1}{{\sharp \mathcal{C}^*}^y}\sum_{\{C_j\}\in{\mathcal{C}^*}^{y}} \bra{\rho_0} \prod_{j=y}^1 C_j^{-1}\Lambda C_j\ket{\rho_0}\\
&\equiv \bra{\rho_0} \Lambda_{\rm{twirl}}^y\ket{\rho_0}\text{.}\label{eqn:fid1}
\end{align}
The matrix $\Lambda_{\rm{twirl}}$ representing the error channel $\Lambda$ twirled over $\mathcal{C}^\ast$ now acting solely on $\mathcal{H}_d$ has only real non-negative entries since they represent the transition from one level to another. The Perron--Frobenius theorem of linear algebra \cite{Perron1907,Frobenius1912,Meyer2000} states that such a matrix always has a unique highest eigenvalue with positive eigenvector under the condition that all entries are nonzero which is clearly given for any real quantum channel. This eigenvector represents the state the system converges to for increasing sequence length $y$, irrespective of the initial state, and the single unique highest eigenvalue has to be one.  Another fundamental property of any \La in this  eigenbasis is that every column sums up to one due to trace preservation. Here, the trace of a vector means the sum of its vector entries as each basis vector represents a density matrix with unit trace.

 $\Lambda_{\rm{twirl}}$ can be assumed to be diagonalizable in all cases besides a set with vanishing Haar measure; its eigenvectors  $\lambda_i$, on the other hand, are not restricted to be real for more than one qubit as the resulting channel is no longer guaranteed to be depolarizing. To investigate conditions for real eigenvalues we first consider a  channel \La on $\mathcal{H}_d$ resulting of a small unitary error. We write $\hat{U}=\exp(-i \Delta \hat{H})$ as an exponential of a Hermitian operator with $\Delta\|\hat{H}\|\ll1$ and expand the resulting $\Lambda$ in orders of a small parameter $\Delta$. Since the zeroth order of the expansion is the identity and the first order vanishes by construction the leading order matrix elements to be investigated are of second order in $\Delta$, 
\begin{align}
 \Lambda_{ij}^{(2)}&=\Delta^2 \text{Tr}\left[\ketbra{i}{i}\hat{H}\ketbra{j}{j}\hat{H}\right]\notag\\
&=\Delta^2 |\bra{i}\hat{H}\ket{j}|^2=\Lambda_{ji}^{(2)}\text{.}
\end{align}
\La is thus Hermitian up to second order ensuring real eigenvalues $\lambda_i$. Third order terms have small effects on the characteristic polynomial and of \La and complex eigenvalues can only occur if the second order eigenvalues are degenerate or very close to each other. For a most general \La one has to consider an unitary on the system combined with an environment $\{\ket{E_k}\}$ \cite{Nielsen2000}.
\begin{align}
\Lambda_{ij}^{(2)}&=\Delta^2\sum_k|\bra{i,E_k}\hat{H}\ket{j,E_0}|^2
\end{align}
Only one summand, $k=0$, is guaranteed to be invariant under exchange of $i$ and $j$, the rest are arbitrary and can therefore in principle yield eigenvalues $\lambda_i$ which have a non-negligible imaginary part. It needs to be noted that this is not very common as these terms are typically close to Hermitian operators, as is also confirmed numerically. 
In any case $\Lambda_{\rm{twirl}}^y$ is an element of the Hilbert space  of linear operators on $\mathcal{H}_d$, and  $\Phi_y$ of equation (\ref{eqn:fid1}) is a linear functional of that, therefore 
\begin{align}
\Phi_y=\sum_ia_i\lambda_i^y
\end{align}
with constants $a_i$ defining the functional in the eigenbasis of $\Lambda_{\rm{twirl}}$. Note that this is the same form as equation (\ref{eqn:model}). Unlike the single qubit case \cite{Epstein2014} for multiple qubits not all errors channels have the same eigenbasis so the fit parameters do not reveal all channel properties of the twirled error. Luckily one does not need to know the exact form of \La but can simply fit the measured fidelity to the above fit model. In the same way it is possible to measure the population of the computational subspace within the same model to characterize the total leakage rate of an error channel. This has  been studied Ref. in \cite{Wallman2014X} under the assumption of control over the leakage space, whereas our work shows that such control is not required for extracting the fidelity alone. 

Fitting a multi-exponential decay with complex decay parameters $\lambda_i$ may pose a problem since oscillations in $\Phi_y$, as they are overlapping with each other and the natural fluctuations, can be hard to fit especially when small. To make a safe estimation of the maximum average gate error one has to consider that oscillations in fidelity are population coming back into the initial state. Since we twirl over the computational subspace the maximum factor for repopulating this state is $\frac{1}{2^N}$ so the fitted $\Phi_y$ acknowledges at least a fraction of $1-2^{-N}$ of the actual error. 

\subsection{RB of Clifford sets that are not twirl designs}
\label{subsec:unlike}
The assumption made in earlier this section when introducing the twirl design that the entangling gate or gates are diagonal on the non computational subspace does not apply to all physical systems. This assumption was necessary to make sure the randomized Clifford set closes like a group hence forming a twirl design. For example considering a CNOT on the computational subspace also performing a NOT on the target if the control qubit is in $\ket{2}$. Omitting this assumption thus enforces to reconsider equation (\ref{eqn:fid0}) where $\mathcal{C}_{y+1}$ is no longer the inverse of all previous Clifford gates as the dynamics inside the non-computational space are not accounted for and may be non-Clifford. Because we are only measuring within the computational subspace -- up to measurement error that we neglect as it contributes to SPAM -- we can calculate as it were 
\begin{align}
\Phi_y&= \frac{1}{{\sharp \mathcal{C}^*}^y}\sum_{\{C_j\}\in{\mathcal{C}^*}^{y}} \bra{\rho_0} C_1^{-1}C_1C_{y+1}\prod_{j=y}^2 (\Lambda C_j) \Lambda C_1\ket{\rho_0}\text{.}
\end{align}
With $C_1C_{y+1}$ being the inverse of all Cliffords but the first, one can see that applying one more random Clifford gate in the sequence means multiplying $C_1^{-1}$ from the left and $\Lambda C_1$ from the right averaged over all $C_1$ in $\mathcal{C}^*$. Thus, extending the length of the sequence by one gate defines a linear and positive map $T_\Lambda$ on the Hilbert space of linear operators on $\mathcal{H}_d$. The resulting fidelity
\begin{align}
\Phi_y&= \bra{\rho_0} T_\Lambda^y(\bbone)\ket{\rho_0}\
\end{align}
as a linear functional of a matrix exponential is again
\begin{align}
\Phi_y&= \sum_i b_i \kappa_i^y\text{.}
\end{align}
There are vanishing entries due to the fact that the Clifford set does not connect all states in particular it does not connect leakage and computational subspace. According to the Perron--Frobenius theorem the highest eigenvalue does no longer have to be unique but the moduli of all eigenvalues are still bound by one. This implies that there might no longer be a uniform limit to which all elements of $\mathcal{H}_d$ converge for large $y$. It is but true that the protocol is still invariant under changing the initial state in the computational subspace, so only accounting for real eigenvalues produces a factor of $1-2^{-N}$ which vanishes for long sequences. 

\begin{figure}[htbp!]
 \centering
 \includegraphics[width=0.45\textwidth]{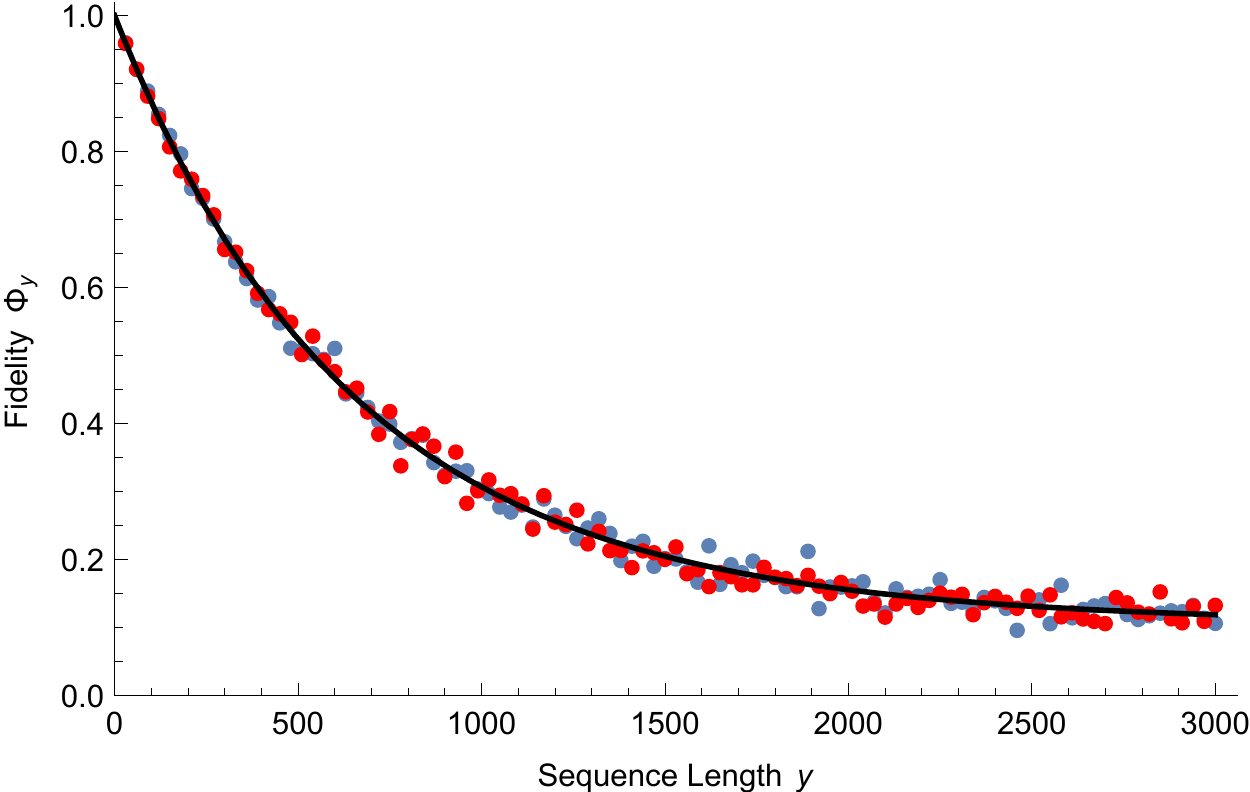}
 \caption{Average fidelity $\Phi_y$ for a unitary error of $1.354 \times 10^{-3}$ for 40 different random sequences per sequence length $y$. The blue points are simulated data for a twirl design whilst the Clifford set used to generate the red points is none. The twirl design estimates an error of $1.379 \times 10^{-3}$ and the other $1.350 \times 10^{-3}$.  
Both 2-qubit Clifford sets are generated from the single qubit Clifford sets and few 2-qubit entangling gates \cite{Corcoles2013} which do or do not satisfy the conditions for a twirl design. The obtained average sequence fidelities $\Phi_y$ are fitted to the multi exponential decay function  and the error rates are calculated via equation (\ref{eqn:rfid}). }
 \label{fig:bad}
\end{figure}

Figure \ref{fig:bad} compares the RB protocol for examples of Clifford sets which are or are not twirl designs for a unitary error which as the channel most affected by twirling and therefore has been considered the errors most difficult to estimate with RB in previous publications. However the actual difference in average fidelity is small. This is due to the fact that the twirl on the computational subspace is intact and that the Clifford set is not changed much by shifting of one of its elements.

We produce a random unitary error by applying a random unitary basis transformation to the exponential of a small diagonal Hermitian. For arbitrary error channels we apply such a unitary to the system plus an environment which we then trace out.

\subsection{State Preparation and Measurement Errors -- SPAM}
\label{subsec:spam}
In all the above calculations we neglect SPAM errors. We show in this subsection that this is justified and that SPAM errors have in fact very little influence on our RB protocol and can moreover, as in other RB protocols, be isolated as they do not depend on the sequence length $y$ thus giving an offset of the fidelity curve. 
For twirl designs, none of the calculations assume that one prepares and measures states in the computational subspace. 
Due to the phase randomization SPAM errors can be assumed to act on $\mathcal{H}_d$ leading to the following fidelity of the full experiment,
\begin{align}
\Phi_y&=\bra{\rho_0}\Lambda_M\Lambda \Lambda_{\rm{twirl}}^y \Lambda_P\ket{\rho_0}\notag\\
&\equiv \bra{\rho_M}\Lambda_{\rm{twirl}}^y \ket{\rho_P}.
\end{align}
Here,  the second line defines the actually prepared state $\rho_P$, as well as an effective measured state $\rho_M$.
As this is once again a linear functional of a matrix exponential the model stays intact. One can think of the actual average fidelity of a single random Clifford as the factor by which the highest possible effective fidelity  decreases due to its application. Said fidelity is determined by SPAM errors leading to an average fidelity of the Clifford set
\begin{align}
\Phi&=\frac{\bra{\rho_M}\Lambda_{\rm{twirl}} \ket{\rho_P}}{\braket{\rho_M|\rho_P}}=\frac{\Phi_1}{\Phi_0}\label{eqn:rfid}\text{.}
\end{align}
The assumption that for Clifford sets which are not twirl designs the last gate $C_{y+1}$ can be considered the inverse of all previous ones is only true in the computational subspace, because the full operations are not necessarily Clifford. The error made by that is calculated by first splitting up the effective measured state into computational and non-computational parts
\begin{align}
 \Phi_y&= \frac{1}{{\sharp \mathcal{C}^*}^y}\sum_{\{C_j\}\in{\mathcal{C}^*}^{y}} \bra{\rho_M} C_{y+1}\prod_{j=y}^1 (\Lambda C_j) \ket{\rho_P}\notag\\
&=\frac{1}{{\sharp \mathcal{C}^*}^y}\sum_{\{C_j\}\in{\mathcal{C}^*}^{y}} \left[\bra{\rho_M}_{\rm{comp}} C_{y+1}\prod_{j=y}^1 (\Lambda C_j) \ket{\rho_P}\right. \notag\\
&\hspace{1cm}\left. +\bra{\rho_M}_{\rm{leak}} C_{y+1}\prod_{j=y}^1 (\Lambda C_j) \ket{\rho_P}\right]\notag\\
\intertext{For the computational subspace $C_{y+1}$ can now be replaced by the actual inverse.}
&=\frac{1}{{\sharp \mathcal{C}^*}^y}\sum_{\{C_j\}\in{\mathcal{C}^*}^{y}} \left[\bra{\rho_M}_{\rm{comp}} D_y^{-1}\prod_{j=y}^1 (\Lambda C_j) \ket{\rho_P}\right. \notag\\
&\hspace{1cm}+\bra{\rho_M}_{\rm{leak}} \prod_{j=y+1}^1 (C_j\Lambda) C_1\ket{\rho_P}_{\rm{comp}}\notag\\
&\hspace{1cm}\left. +\bra{\rho_M}_{\rm{leak}} C_{y+1}\prod_{j=y}^1 (\Lambda C_j)\ket{\rho_P}_{\rm{leak}}\right]\notag\\
&=\bra{\rho_M}_{\rm{comp}} T_\Lambda^y(\bbone) \ket{\rho_P}\notag\\
&+\bra{\rho_M}_{\rm{leak}} \tilde{T}_\Lambda^y(\bbone)\ket{\rho_P}_{\rm{comp}}+\mathcal{O}(\varepsilon_P(\varepsilon_M+\varepsilon)).
\end{align}
The $\varepsilon$ are  state preparation, measurement and gate error rates  and hence the deviation from the model is negligibly small and not scaling with $y$ leading to the same fidelity arguments as before.
\section{Variance of the Fidelity}
\label{sec:var}
The number of possible choices for $y$ random Clifford gates is $\left(\sharp \mathcal{C}_N~2^N\right)^y$ where the cardinality of the Clifford group has a lower bound exponential in N. With that background it is remarkable how few different realizations of sequences are typically needed to obtain a reliable result. We therefore investigate the variance of the fidelity with respect to these possible choices up to second/third order  for twirl designs. (Although all error are considered the same they are perturbed 
by $\Lambda=\Lambda_{\rm{twirl}}+\epsilon_\Lambda$ in both factors of the squared term as well as after every gate individually.) The raw channel \La is close to $\Lambda_{\rm twirl}$ but has an error due to the limited number of sequences, and therefore can be written as $\Lambda=\Lambda_{\rm{twirl}}+\epsilon_\Lambda$ with a small operator $\epsilon_\Lambda$. Sorting by order of this $\epsilon_\Lambda$ and omitting higher orders one gets
\begin{align}
&[\Phi^2]_{\mathcal{C}^*}=[\Phi]_{\mathcal{C}^*}^2+\notag{}\\
&\frac{1}{\sharp \mathcal{C}^*}\sum_{x=1}^y\sum_{D\in\mathcal{C}^*}\bra{\rho_0}\Lambda_{\rm{twirl}}^{x-1}D^\dagger\epsilon_\Lambda D \Lambda_{\rm{twirl}}^{y-x}\ket{\rho_0}^2 \label{eqn:error channel} \text{.}
\end{align}
Note that the above is second order since every twirl over an $\epsilon_\Lambda$ has to vanish. Also all second order terms with perturbations on different errors are zero as well as those of every odd order because single deviation always twirls to zero. Furthermore one needs to take into account that the errors  $\epsilon_\Lambda$ as the difference between an operator on $\mathcal{H}_d$ and one on density matrices in general is not per se an operator on $\mathcal{H}_d$ but rather acts on the space of density operators. Since for the second order terms there is only one $\epsilon_\Lambda$ in each $\rho_0$ bracket it can be just as well projected onto such. For further calculations we change to the eigenbasis of the twirled error and abbreviate $\epsilon(D)\equiv D^\dagger\epsilon_\Lambda D$. We can thus work out the RHS of equation (\ref{eqn:error channel}).
\begin{align}
&\frac{1}{\sharp \mathcal{C}^*}\sum_{x=1}^y\sum_{D\in\mathcal{C}^*}\bra{\rho_0}\Lambda_{\rm{twirl}}^{x-1}D^\dagger\epsilon_\Lambda D \Lambda_{\rm{twirl}}^{y-x}\ket{\rho_0}^2\notag\\
&=\frac{1}{\sharp \mathcal{C}^*}\sum_{x=1}^y\sum_{D\in\mathcal{C}^*}\sum_{s_1,s_2,s_3,s_4}^{3^N}\notag\\
&\hspace{1cm} \gamma_{\{s\}}\lambda_{s_1}^{x-1}\epsilon(D)_{s_1s_2}\lambda_{s_2}^{y-x}\lambda_{s_3}^{x-1}\epsilon(D)_{s_3s_4}\lambda_{s_4}^{y-x}\notag\\
&=\frac{1}{\sharp \mathcal{C}^*}\sum_{D,\{s\}}\sum_{x=1}^y\notag\\
&\hspace{1cm}\gamma_{\{s\}}\left(\frac{\lambda_{s_1}\lambda_{s_3}}{\lambda_{s_2}\lambda_{s_4}}\right)^x\frac{\left(\lambda_{s_2}\lambda_{s_4}\right)^y}{\lambda_{s_1}\lambda_{s_3}}\epsilon(D)_{s_1s_2}\epsilon(D)_{s_3s_4}\label{eqn:easyvar}\\
\intertext{Evaluating the geometric sum leaves us with}
&\equiv\sum_{\{s\}}\gamma_{\{s\}}\frac{1-\left(\frac{\lambda_{s_1}\lambda_{s_3}}{\lambda_{s_2}\lambda_{s_4}}\right)^{y+1}}{1-\left(\frac{\lambda_{s_1}\lambda_{s_3}}{\lambda_{s_2}\lambda_{s_4}}\right)}
\frac{\left(\lambda_{s_2}\lambda_{s_4}\right)^y}{\lambda_{s_1}\lambda_{s_3}}E_{s_1s_2}E_{s_3s_4}\text{.}\label{eqn:var}
\end{align}
\begin{figure}[htbp!]
 \centering
 \includegraphics[width=0.45\textwidth]{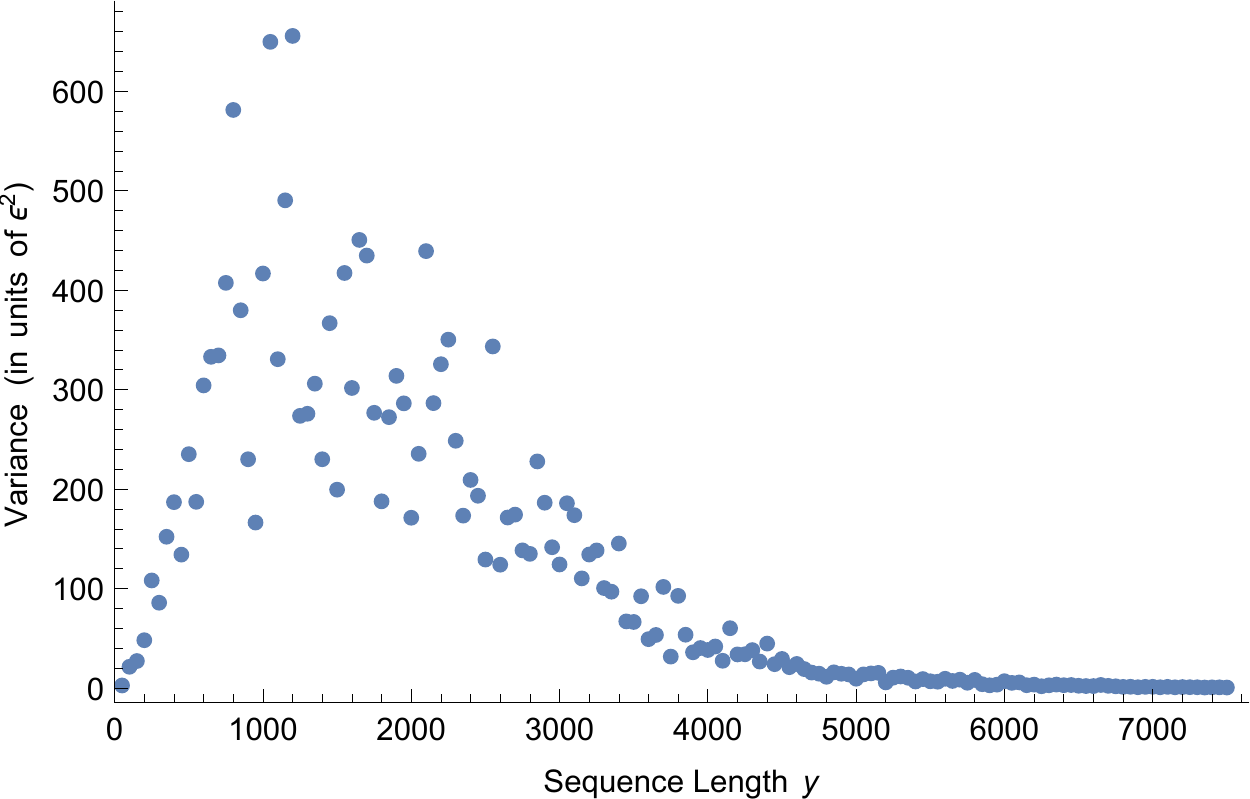}
 \caption{Variance with respect to possible combinations of Clifford gates for an relatively small error ($7.42 \times 10^{-4}$) -- averaged over a total of 30 different sequences}
 \label{fig:var}
\end{figure}
Equation (\ref{eqn:var}) always has an upper bound by an exponential decay times $y$, $\gamma$ is smaller or in the order of magnitude than/of one and the $E$s are comparable to single gate errors. For simplicity (and consistent with our numerics these terms are mostly bigger than others for random errors) we now assume that $s_1=s_2$ and $s_3=s_4$ and continue on equation (\ref{eqn:easyvar})
\begin{align}
\left([\Phi]_{\mathcal{C}^*}^2-[\Phi^2]_{\mathcal{C}^*}\right)_{s_1,s_3}&=\gamma_{\{s\}}y\frac{\left(\lambda_{s_1}\lambda_{s_3}\right)^y}{\lambda_{s_1}\lambda_{s_3}}E_{s_1s_1}E_{s_3s_3}\\
&\propto ye^{-\kappa y}
\end{align}
This explains the behavior observed in the simulation conducted for figure \ref{fig:var}. For $s_1=1$ or $s_3=1$ which corresponds to changes of the state with $\lambda=1$ the E terms vanish since this state is preserved for all quantum channels. Therefore there will always be an exponential decay. The order of this estimates is the same as that for standard RB \cite{Wallman2014}.
\section{Gate dependent Errors}
\label{sec:gdep}
The RB protocol analyzed so far assumes \La to be independent of the preceding gate or more specifically uses an effective gate error. Magesan \emph{et al.} investigated first order corrections to that assumption for the leakage free case yielding an altered fidelity model under the condition $\|\Lambda-\Lambda_j\|y\ll1$ with gate dependent error channels $\Lambda_j$ \cite{Magesan2011}. Having more degrees of freedom for these error channels $\Lambda_j$ this issue becomes even more important for the leakage case. Neglecting SPAM we get
\begin{align}
\Phi_y&= \frac{1}{{\sharp \mathcal{C}^*}^y}\sum_{\{C_j\}\in{\mathcal{C}^*}^{y}} \bra{\rho_0}\Lambda_{y+1} C_{y+1}\prod_{j=y}^1 (\Lambda_j C_j) \ket{\rho_0}\notag\\
&\cong \frac{1}{{\sharp \mathcal{C}^*}^y}\sum_{\{C_j\}\in{\mathcal{C}^*}^{y}} \bra{\rho_0}\Lambda C_{y+1}\prod_{j=y}^1 (\Lambda_j C_j) \ket{\rho_0}\notag\\
&=\frac{1}{{\sharp \mathcal{C}^*}^y}\sum_{\{C_j\}\in{\mathcal{C}^*}^{y}} \bra{\rho_0}\Lambda C_1^{-1}~C_1C_{y+1}~\times\notag\\
&\hspace{2cm}\times~\prod_{j=y}^2 (\Lambda_j C_j)~\Lambda_1C_1 \ket{\rho_0}\text{.}
\end{align}
Up to an additional error of the order of $\|\Lambda-\Lambda_j\|y\ll1$  which is overall and not per gate, the expansion by one random Clifford is a linear operation on the operators on $\mathcal{H}_d$
hence
\begin{align}
 \Phi_y&=\bra{\rho_0} T_G^y(\bbone)\ket{\rho_0}\\
&=\sum_ic_i\tau_i^y\text{.}
\end{align}
This shows that is it not necessary to alter the protocol for gate dependent errors; however there a two aspects which need to be considered.\\
\textit{First} the phase randomization is no longer considered perfect leading to transitions to non diagonal density matrices. This has the potential of blowing up the dimension of $T_G$ up by a factor of $3^{2N}$ resulting in more and likely complex eigenvalues but as second order terms of single qubit errors these effects can be considered small with respect to the average error. In the practical application one needs to consider how many different $\tau_i$ are needed to properly describe the $y$ dependence of the fidelity. One can see that the number of different $\tau_i$ is  the order up to which deviation from an effective $\tau$ are completely covered. Magesan \emph{et al.} discussed conditions for neglecting higher order terms when they derived a first order protocol for gate dependent errors without leakage \cite{Magesan2011}. As our calculation still holds without leakage errors one can show how their model coincides with ours in the limit of sufficiently weak gate dependencies while we are not restricted to this regimes.\\
\textit{Second} we have to note that $\Phi_y$ is no longer completely independent of the initial state which is necessary for considering only real eigenvalues and counteracts the idea of RB itself. It however can be fixed by an approximation making a single error of the order of $\|\Lambda-\Lambda_j\|$
\begin{align}
 \Phi_y&=\frac{1}{{\sharp \mathcal{C}^*}}\sum_{C\in{\mathcal{C}^*}}\bra{\rho_0}C^{-1} T_G^{y-1}(\bbone)\Lambda C\ket{\rho_0}\text{.}
\end{align}
This means that the protocol is independent of the initial as well as the measured state up to a very small error. This aspect is crucial for that fitting to real eigenvalues accounts for at least $1-2^{-N}$ of the actual error. The gate dependent error channels produce a fundamentally different error associated matrix $T$ than in the gate independent case so one can no longer argue that all real eigenvalues are likely, regardless of the use of the phase randomizing Clifford set $\mathcal{C}^\ast$.   

\section{Interleaved Randomized Benchmarking}
\label{sec:irb}
The RB protocol is designed to identify the average performance of a unitary 2-design such as the Clifford group; the fidelity of an individual element $V$ thereof can be estimated by Interleaved Randomized Benchmarking (IRB) \cite{Magesan2012}. The main idea is to alternate random Clifford gates with gate of interest to obtain a combined fidelity. Applied to our framework one obtains
\begin{align}
\Phi_y&= \frac{1}{{\sharp \mathcal{C}^*}^y}\sum_{\{C_j\}\in{\mathcal{C}^*}^{y}} \bra{\rho_0}\Lambda_{y+1} C_{y+1}\prod_{j=y}^1 (V \Lambda_V \Lambda_j C_j) \ket{\rho_0}\notag\\
&\cong \frac{1}{{\sharp \mathcal{C}^*}^y}\sum_{\{C_j\}\in{\mathcal{C}^*}^{y}} \bra{\rho_0}\Lambda C_1^{-1}V^{-1} VC_1C_{y+1}~\times\notag\\
&\hspace{1.5cm}\times~\prod_{j=y}^2 (V \Lambda_V \Lambda_j C_j) V\Lambda_V \Lambda_1 C_1 \ket{\rho_0}\text{.}
\end{align}
$VC_1C_{y+1}$ is the inverse of the gate sequence without $VC_1$ and therefore adding a random gate plus the gate in question is once again a linear operation which results in the same fit model
\begin{align}
  \Phi_y&=\bra{\rho_0} T_V^y(\bbone)\ket{\rho_0}\\
&=\sum_id_i\eta_i^y\text{.}
\end{align}
The resulting fidelity
\begin{align}
 \Phi_1&=\frac{1}{{\sharp \mathcal{C}^*}}\sum_{C\in{\mathcal{C}^*}}\bra{\rho_0} C^{-1}V^{-1}V\Lambda_V\Lambda_CC\ket{\rho_0}\notag\\
&=\frac{1}{{\sharp \mathcal{C}^*}}\sum_{C\in{\mathcal{C}^*}}\bra{\rho_0} C^{-1}\Lambda_V\Lambda_CC\ket{\rho_0}
\end{align}
shows the average combined error of $V$ and the Clifford set.  Assuming those are small the estimated error rate of $V$ can be calculated as $\varepsilon_V=\varepsilon_{\mathcal{C}^\ast \times V}-\varepsilon_{\mathcal{C}^\ast}$ and lies within the bounds
\begin{align} 
\left(\sqrt{\varepsilon_{\mathcal{C}^\ast \times V}}-\sqrt{\varepsilon_{\mathcal{C}^\ast}}\right)^2\leq \varepsilon_V\leq\left(\sqrt{\varepsilon_{\mathcal{C}^\ast \times V}}+\sqrt{\varepsilon_{\mathcal{C}^\ast}}\right)^2\text{.}
\end{align}

\begin{figure}[htbp!]
 \centering
 \includegraphics[width=0.45\textwidth]{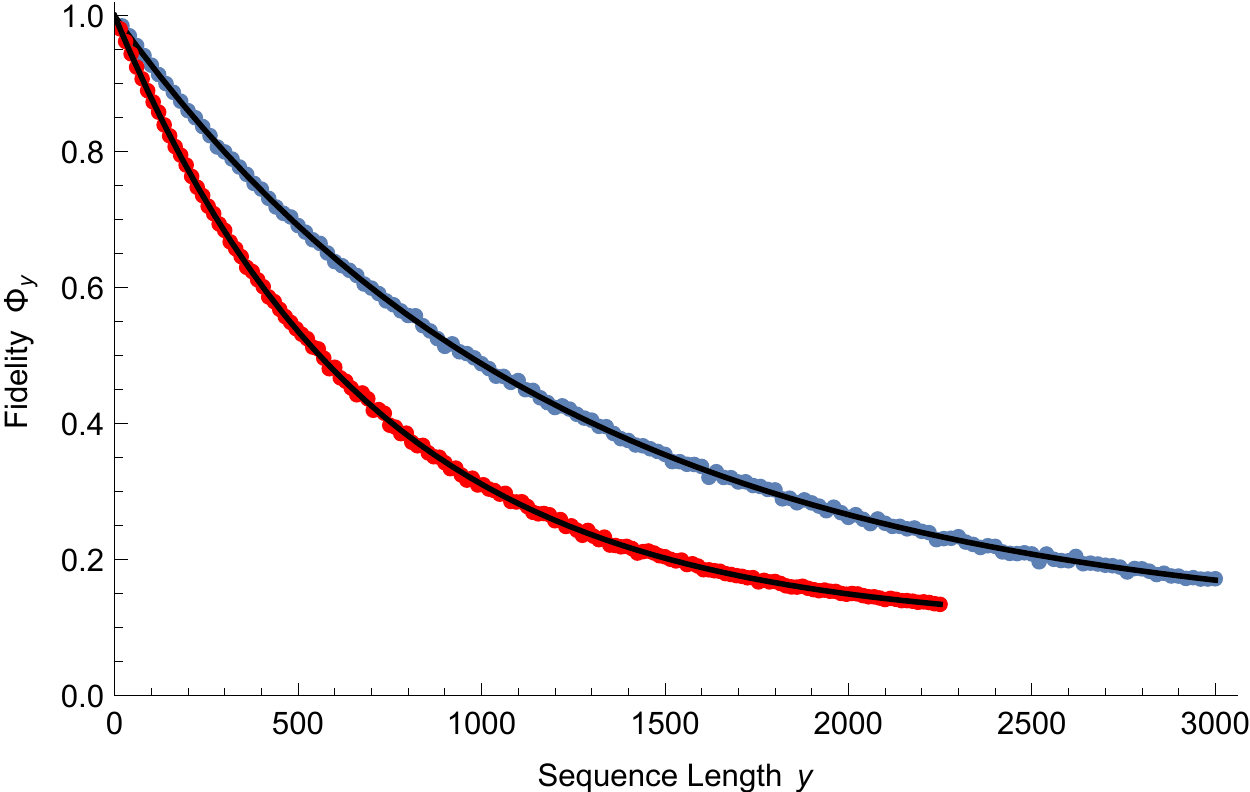}
 \caption{Interleaved Randomized Benchmarking for an unitary operation on $\mathcal{SU}(4)$ not element of a twirl design with a gate independent error of $7.57 \times 10^{-4}$ (blue). The interleaved protocol yields a combined error of $1.304 \times 10^{-3}$ estimating the gate of interest at $5.50 \times 10^{-4}$ when its actual error is $5.44 \times 10^{-4}$. This is a very good estimate considering the range of IRB.}
 \label{fig:int}
\end{figure}

The only point where we used that $V$ is in $\mathcal{C}^*$ is that the inverting gate at the end of the sequence can be implemented sufficiently accurately. This means that its error has to be small with respect to the sequence errors but can easily be of an order of magnitude higher than the gate error because it is independent of the sequence length. This is comparable to the influence of SPAM errors. Under above assumption the IRB protocol holds for every unitary operation which is shown in figure \ref{fig:int}.
\section{Conclusion}
We have investigated the effect of leakage errors on RB and, to account for those, provided an alternated protocol using a phase randomizing Clifford set and a multi-exponential decay fit function. We introduced conditions under which the twirl of the error channel over the Clifford set is preserved but also showed that our protocol applies regardless of this. Although we only considered one additional level out work is easily expendable to multiple leakage levels. Our protocol accounts for arbitrary gate dependent error channels, even if they are not small perturbations. We showed that our protocol is compatible with IRB and that this even holds for gates that are not element of the underlying unitary 2-design.
\section{Acknowledgements}
We acknowledge useful discussions with Rami Barends and Jimmy Chen. This work was supported by the EU through SCALEQIT and QUAINT and funded by the Office of the Director of National Intelligence (ODNI), Intelligence Advanced Research Projects Activity (IARPA), through the Army Research Office. All statements of fact, opinion or conclusions contained herein are those of the authors and should not be construed as representing the official views or policies of IARPA, the ODNI, or the U.S. Government.

\bibliography{LRB.bib}

\begin{thebibliography}{23}%
\makeatletter
\providecommand \@ifxundefined [1]{%
 \@ifx{#1\undefined}
}%
\providecommand \@ifnum [1]{%
 \ifnum #1\expandafter \@firstoftwo
 \else \expandafter \@secondoftwo
 \fi
}%
\providecommand \@ifx [1]{%
 \ifx #1\expandafter \@firstoftwo
 \else \expandafter \@secondoftwo
 \fi
}%
\providecommand \natexlab [1]{#1}%
\providecommand \enquote  [1]{``#1''}%
\providecommand \bibnamefont  [1]{#1}%
\providecommand \bibfnamefont [1]{#1}%
\providecommand \citenamefont [1]{#1}%
\providecommand \href@noop [0]{\@secondoftwo}%
\providecommand \href [0]{\begingroup \@sanitize@url \@href}%
\providecommand \@href[1]{\@@startlink{#1}\@@href}%
\providecommand \@@href[1]{\endgroup#1\@@endlink}%
\providecommand \@sanitize@url [0]{\catcode `\\12\catcode `\$12\catcode
  `\&12\catcode `\#12\catcode `\^12\catcode `\_12\catcode `\%12\relax}%
\providecommand \@@startlink[1]{}%
\providecommand \@@endlink[0]{}%
\providecommand \url  [0]{\begingroup\@sanitize@url \@url }%
\providecommand \@url [1]{\endgroup\@href {#1}{\urlprefix }}%
\providecommand \urlprefix  [0]{URL }%
\providecommand \Eprint [0]{\href }%
\providecommand \doibase [0]{http://dx.doi.org/}%
\providecommand \selectlanguage [0]{\@gobble}%
\providecommand \bibinfo  [0]{\@secondoftwo}%
\providecommand \bibfield  [0]{\@secondoftwo}%
\providecommand \translation [1]{[#1]}%
\providecommand \BibitemOpen [0]{}%
\providecommand \bibitemStop [0]{}%
\providecommand \bibitemNoStop [0]{.\EOS\space}%
\providecommand \EOS [0]{\spacefactor3000\relax}%
\providecommand \BibitemShut  [1]{\csname bibitem#1\endcsname}%
\let\auto@bib@innerbib\@empty
\bibitem [{\citenamefont {Barends}\ \emph {et~al.}(2014)\citenamefont
  {Barends}, \citenamefont {Kelly}, \citenamefont {Megrant}, \citenamefont
  {Veitia}, \citenamefont {Sank}, \citenamefont {Jeffrey}, \citenamefont
  {White}, \citenamefont {Mutus}, \citenamefont {Fowler}, \citenamefont
  {Campbell}, \citenamefont {Chen}, \citenamefont {Chen}, \citenamefont
  {Chiaro}, \citenamefont {Dunsworth}, \citenamefont {Neill}, \citenamefont
  {O’Malley}, \citenamefont {Roushan}, \citenamefont {Vainsencher},
  \citenamefont {Wenner}, \citenamefont {Korotkov}, \citenamefont {Cleland},\
  and\ \citenamefont {Martinis}}]{Barends2014}%
  \BibitemOpen
  \bibfield  {author} {\bibinfo {author} {\bibfnamefont {R.}~\bibnamefont
  {Barends}}, \bibinfo {author} {\bibfnamefont {J.}~\bibnamefont {Kelly}},
  \bibinfo {author} {\bibfnamefont {A.}~\bibnamefont {Megrant}}, \bibinfo
  {author} {\bibfnamefont {A.}~\bibnamefont {Veitia}}, \bibinfo {author}
  {\bibfnamefont {D.}~\bibnamefont {Sank}}, \bibinfo {author} {\bibfnamefont
  {E.}~\bibnamefont {Jeffrey}}, \bibinfo {author} {\bibfnamefont
  {T.}~\bibnamefont {White}}, \bibinfo {author} {\bibfnamefont
  {J.}~\bibnamefont {Mutus}}, \bibinfo {author} {\bibfnamefont
  {A.}~\bibnamefont {Fowler}}, \bibinfo {author} {\bibfnamefont
  {B.}~\bibnamefont {Campbell}}, \bibinfo {author} {\bibfnamefont
  {Y.}~\bibnamefont {Chen}}, \bibinfo {author} {\bibfnamefont {Z.}~\bibnamefont
  {Chen}}, \bibinfo {author} {\bibfnamefont {B.}~\bibnamefont {Chiaro}},
  \bibinfo {author} {\bibfnamefont {A.}~\bibnamefont {Dunsworth}}, \bibinfo
  {author} {\bibfnamefont {C.}~\bibnamefont {Neill}}, \bibinfo {author}
  {\bibfnamefont {P.}~\bibnamefont {O’Malley}}, \bibinfo {author}
  {\bibfnamefont {P.}~\bibnamefont {Roushan}}, \bibinfo {author} {\bibfnamefont
  {A.}~\bibnamefont {Vainsencher}}, \bibinfo {author} {\bibfnamefont
  {J.}~\bibnamefont {Wenner}}, \bibinfo {author} {\bibfnamefont
  {A.}~\bibnamefont {Korotkov}}, \bibinfo {author} {\bibfnamefont
  {A.}~\bibnamefont {Cleland}}, \ and\ \bibinfo {author} {\bibfnamefont
  {J.}~\bibnamefont {Martinis}},\ }\href@noop {} {\bibfield  {journal}
  {\bibinfo  {journal} {Nature}\ }\textbf {\bibinfo {volume} {508}},\ \bibinfo
  {pages} {500} (\bibinfo {year} {2014})}\BibitemShut {NoStop}%
\bibitem [{\citenamefont {Chow}\ \emph {et~al.}(2014)\citenamefont {Chow},
  \citenamefont {Gambetta}, \citenamefont {Magesan}, \citenamefont
  {Srinivasan}, \citenamefont {Cross}, \citenamefont {Abraham}, \citenamefont
  {Masluk}, \citenamefont {Johnson}, \citenamefont {Ryan},\ and\ \citenamefont
  {Steffen}}]{Chow2014}%
  \BibitemOpen
  \bibfield  {author} {\bibinfo {author} {\bibfnamefont {J.}~\bibnamefont
  {Chow}}, \bibinfo {author} {\bibfnamefont {J.}~\bibnamefont {Gambetta}},
  \bibinfo {author} {\bibfnamefont {E.}~\bibnamefont {Magesan}}, \bibinfo
  {author} {\bibfnamefont {S.}~\bibnamefont {Srinivasan}}, \bibinfo {author}
  {\bibfnamefont {A.}~\bibnamefont {Cross}}, \bibinfo {author} {\bibfnamefont
  {D.}~\bibnamefont {Abraham}}, \bibinfo {author} {\bibfnamefont
  {N.}~\bibnamefont {Masluk}}, \bibinfo {author} {\bibfnamefont
  {B.}~\bibnamefont {Johnson}}, \bibinfo {author} {\bibfnamefont
  {C.}~\bibnamefont {Ryan}}, \ and\ \bibinfo {author} {\bibfnamefont
  {M.}~\bibnamefont {Steffen}},\ }\href@noop {} {\bibfield  {journal} {\bibinfo
   {journal} {Nat. Commun.}\ }\textbf {\bibinfo {volume} {5}},\ \bibinfo
  {pages} {4015} (\bibinfo {year} {2014})}\BibitemShut {NoStop}%
\bibitem [{\citenamefont {Harty}\ \emph {et~al.}(2014)\citenamefont {Harty},
  \citenamefont {Allcock}, \citenamefont {Ballance}, \citenamefont {Guidoni},
  \citenamefont {Janacek}, \citenamefont {Linke}, \citenamefont {Stacey},\ and\
  \citenamefont {Lucas}}]{Harty2014}%
  \BibitemOpen
  \bibfield  {author} {\bibinfo {author} {\bibfnamefont {T.}~\bibnamefont
  {Harty}}, \bibinfo {author} {\bibfnamefont {D.}~\bibnamefont {Allcock}},
  \bibinfo {author} {\bibfnamefont {C.}~\bibnamefont {Ballance}}, \bibinfo
  {author} {\bibfnamefont {L.}~\bibnamefont {Guidoni}}, \bibinfo {author}
  {\bibfnamefont {H.}~\bibnamefont {Janacek}}, \bibinfo {author} {\bibfnamefont
  {N.}~\bibnamefont {Linke}}, \bibinfo {author} {\bibfnamefont
  {D.}~\bibnamefont {Stacey}}, \ and\ \bibinfo {author} {\bibfnamefont
  {D.}~\bibnamefont {Lucas}},\ }\href@noop {} {\bibfield  {journal} {\bibinfo
  {journal} {Phys. Rev. Lett.}\ }\textbf {\bibinfo {volume} {113}},\ \bibinfo
  {pages} {220501} (\bibinfo {year} {2014})}\BibitemShut {NoStop}%
\bibitem [{\citenamefont {Nielsen}\ and\ \citenamefont
  {Chuang}(2000)}]{Nielsen2000}%
  \BibitemOpen
  \bibfield  {author} {\bibinfo {author} {\bibfnamefont {M.}~\bibnamefont
  {Nielsen}}\ and\ \bibinfo {author} {\bibfnamefont {I.}~\bibnamefont
  {Chuang}},\ }\href@noop {} {\emph {\bibinfo {title} {Quantum Computation and
  Quantum Information}}}\ (\bibinfo  {publisher} {Cambridge University Press},\
  \bibinfo {year} {2000})\BibitemShut {NoStop}%
\bibitem [{\citenamefont {Knill}\ \emph {et~al.}(2008)\citenamefont {Knill},
  \citenamefont {Leibfried}, \citenamefont {Reichle}, \citenamefont {Britton},
  \citenamefont {Blakestad}, \citenamefont {Jost}, \citenamefont {Langer},
  \citenamefont {Ozeri}, \citenamefont {Seidelin},\ and\ \citenamefont
  {Wineland}}]{Knill2008}%
  \BibitemOpen
  \bibfield  {author} {\bibinfo {author} {\bibfnamefont {E.}~\bibnamefont
  {Knill}}, \bibinfo {author} {\bibfnamefont {D.}~\bibnamefont {Leibfried}},
  \bibinfo {author} {\bibfnamefont {R.}~\bibnamefont {Reichle}}, \bibinfo
  {author} {\bibfnamefont {J.}~\bibnamefont {Britton}}, \bibinfo {author}
  {\bibfnamefont {R.}~\bibnamefont {Blakestad}}, \bibinfo {author}
  {\bibfnamefont {J.}~\bibnamefont {Jost}}, \bibinfo {author} {\bibfnamefont
  {C.}~\bibnamefont {Langer}}, \bibinfo {author} {\bibfnamefont
  {R.}~\bibnamefont {Ozeri}}, \bibinfo {author} {\bibfnamefont
  {S.}~\bibnamefont {Seidelin}}, \ and\ \bibinfo {author} {\bibfnamefont
  {D.}~\bibnamefont {Wineland}},\ }\href@noop {} {\bibfield  {journal}
  {\bibinfo  {journal} {Phys. Rev. A}\ }\textbf {\bibinfo {volume} {77}},\
  \bibinfo {pages} {012307} (\bibinfo {year} {2008})}\BibitemShut {NoStop}%
\bibitem [{\citenamefont {Magesan}\ \emph {et~al.}(2011)\citenamefont
  {Magesan}, \citenamefont {Gambetta},\ and\ \citenamefont
  {Emerson}}]{Magesan2011}%
  \BibitemOpen
  \bibfield  {author} {\bibinfo {author} {\bibfnamefont {E.}~\bibnamefont
  {Magesan}}, \bibinfo {author} {\bibfnamefont {J.}~\bibnamefont {Gambetta}}, \
  and\ \bibinfo {author} {\bibfnamefont {J.}~\bibnamefont {Emerson}},\
  }\href@noop {} {\bibfield  {journal} {\bibinfo  {journal} {Phys. Rev. Lett.}\
  }\textbf {\bibinfo {volume} {106}},\ \bibinfo {pages} {180504} (\bibinfo
  {year} {2011})}\BibitemShut {NoStop}%
\bibitem [{\citenamefont {Magesan}\ \emph {et~al.}(2012)\citenamefont
  {Magesan}, \citenamefont {Gambetta}, \citenamefont {Johnson}, \citenamefont
  {Ryan}, \citenamefont {Chow}, \citenamefont {Merkel}, \citenamefont
  {da~Silva}, \citenamefont {Keefe}, \citenamefont {Rothwell}, \citenamefont
  {Ohki}, \citenamefont {Ketchen},\ and\ \citenamefont
  {Steffen}}]{Magesan2012}%
  \BibitemOpen
  \bibfield  {author} {\bibinfo {author} {\bibfnamefont {E.}~\bibnamefont
  {Magesan}}, \bibinfo {author} {\bibfnamefont {J.}~\bibnamefont {Gambetta}},
  \bibinfo {author} {\bibfnamefont {B.~R.}\ \bibnamefont {Johnson}}, \bibinfo
  {author} {\bibfnamefont {C.}~\bibnamefont {Ryan}}, \bibinfo {author}
  {\bibfnamefont {J.}~\bibnamefont {Chow}}, \bibinfo {author} {\bibfnamefont
  {S.}~\bibnamefont {Merkel}}, \bibinfo {author} {\bibfnamefont
  {M.}~\bibnamefont {da~Silva}}, \bibinfo {author} {\bibfnamefont
  {G.}~\bibnamefont {Keefe}}, \bibinfo {author} {\bibfnamefont
  {M.}~\bibnamefont {Rothwell}}, \bibinfo {author} {\bibfnamefont
  {T.}~\bibnamefont {Ohki}}, \bibinfo {author} {\bibfnamefont {M.}~\bibnamefont
  {Ketchen}}, \ and\ \bibinfo {author} {\bibfnamefont {M.}~\bibnamefont
  {Steffen}},\ }\href@noop {} {\bibfield  {journal} {\bibinfo  {journal} {Phys.
  Rev. Lett.}\ }\textbf {\bibinfo {volume} {109}},\ \bibinfo {pages} {080505}
  (\bibinfo {year} {2012})}\BibitemShut {NoStop}%
\bibitem [{\citenamefont {Egger}\ and\ \citenamefont
  {F.K.Wilhelm}(2014)}]{Egger2014}%
  \BibitemOpen
  \bibfield  {author} {\bibinfo {author} {\bibfnamefont {D.}~\bibnamefont
  {Egger}}\ and\ \bibinfo {author} {\bibnamefont {F.K.Wilhelm}},\ }\href@noop
  {} {\bibfield  {journal} {\bibinfo  {journal} {Phys. Rev. Lett.}\ }\textbf
  {\bibinfo {volume} {112}},\ \bibinfo {pages} {240503} (\bibinfo {year}
  {2014})}\BibitemShut {NoStop}%
\bibitem [{\citenamefont {Kelly}\ \emph {et~al.}(2014)\citenamefont {Kelly},
  \citenamefont {Barends}, \citenamefont {Campbell}, \citenamefont {Chen},
  \citenamefont {Chen}, \citenamefont {Chiaro}, \citenamefont {Dunsworth},
  \citenamefont {Fowler}, \citenamefont {Hoi}, \citenamefont {Jeffrey},
  \citenamefont {Megrant}, \citenamefont {Mutus}, \citenamefont {Neill},
  \citenamefont {O’Malley}, \citenamefont {Quintana}, \citenamefont
  {Roushan}, \citenamefont {Vainsencher}, \citenamefont {Wenner}, \citenamefont
  {White}, \citenamefont {Cleland},\ and\ \citenamefont
  {Martinis}}]{Kelly2014}%
  \BibitemOpen
  \bibfield  {author} {\bibinfo {author} {\bibfnamefont {J.}~\bibnamefont
  {Kelly}}, \bibinfo {author} {\bibfnamefont {R.}~\bibnamefont {Barends}},
  \bibinfo {author} {\bibfnamefont {B.}~\bibnamefont {Campbell}}, \bibinfo
  {author} {\bibfnamefont {Y.}~\bibnamefont {Chen}}, \bibinfo {author}
  {\bibfnamefont {Z.}~\bibnamefont {Chen}}, \bibinfo {author} {\bibfnamefont
  {B.}~\bibnamefont {Chiaro}}, \bibinfo {author} {\bibfnamefont
  {A.}~\bibnamefont {Dunsworth}}, \bibinfo {author} {\bibfnamefont
  {A.}~\bibnamefont {Fowler}}, \bibinfo {author} {\bibfnamefont
  {I.}~\bibnamefont {Hoi}}, \bibinfo {author} {\bibfnamefont {E.}~\bibnamefont
  {Jeffrey}}, \bibinfo {author} {\bibfnamefont {A.}~\bibnamefont {Megrant}},
  \bibinfo {author} {\bibfnamefont {J.}~\bibnamefont {Mutus}}, \bibinfo
  {author} {\bibfnamefont {C.}~\bibnamefont {Neill}}, \bibinfo {author}
  {\bibfnamefont {P.}~\bibnamefont {O’Malley}}, \bibinfo {author}
  {\bibfnamefont {C.}~\bibnamefont {Quintana}}, \bibinfo {author}
  {\bibfnamefont {P.}~\bibnamefont {Roushan}}, \bibinfo {author} {\bibfnamefont
  {D.~S.~A.}\ \bibnamefont {Vainsencher}}, \bibinfo {author} {\bibfnamefont
  {J.}~\bibnamefont {Wenner}}, \bibinfo {author} {\bibfnamefont
  {T.}~\bibnamefont {White}}, \bibinfo {author} {\bibfnamefont
  {A.}~\bibnamefont {Cleland}}, \ and\ \bibinfo {author} {\bibfnamefont
  {J.}~\bibnamefont {Martinis}},\ }\href@noop {} {\bibfield  {journal}
  {\bibinfo  {journal} {Phys. Rev. Lett.}\ }\textbf {\bibinfo {volume} {112}},\
  \bibinfo {pages} {240504} (\bibinfo {year} {2014})}\BibitemShut {NoStop}%
\bibitem [{\citenamefont {Gambetta}\ \emph {et~al.}(2012)\citenamefont
  {Gambetta}, \citenamefont {C\'orcoles}, \citenamefont {Merkel}, \citenamefont
  {Johnson}, \citenamefont {Smolin}, \citenamefont {Chow}, \citenamefont
  {Ryan}, \citenamefont {Rigetti}, \citenamefont {Poletto}, \citenamefont
  {Ohki}, \citenamefont {Ketchen},\ and\ \citenamefont
  {Steffen}}]{Gambetta2012}%
  \BibitemOpen
  \bibfield  {author} {\bibinfo {author} {\bibfnamefont {J.}~\bibnamefont
  {Gambetta}}, \bibinfo {author} {\bibfnamefont {A.}~\bibnamefont
  {C\'orcoles}}, \bibinfo {author} {\bibfnamefont {S.}~\bibnamefont {Merkel}},
  \bibinfo {author} {\bibfnamefont {B.}~\bibnamefont {Johnson}}, \bibinfo
  {author} {\bibfnamefont {J.}~\bibnamefont {Smolin}}, \bibinfo {author}
  {\bibfnamefont {J.}~\bibnamefont {Chow}}, \bibinfo {author} {\bibfnamefont
  {C.}~\bibnamefont {Ryan}}, \bibinfo {author} {\bibfnamefont {C.}~\bibnamefont
  {Rigetti}}, \bibinfo {author} {\bibfnamefont {S.}~\bibnamefont {Poletto}},
  \bibinfo {author} {\bibfnamefont {T.}~\bibnamefont {Ohki}}, \bibinfo {author}
  {\bibfnamefont {M.}~\bibnamefont {Ketchen}}, \ and\ \bibinfo {author}
  {\bibfnamefont {M.}~\bibnamefont {Steffen}},\ }\href@noop {} {\bibfield
  {journal} {\bibinfo  {journal} {Phys. Rev. Lett.}\ }\textbf {\bibinfo
  {volume} {109}},\ \bibinfo {pages} {240504} (\bibinfo {year}
  {2012})}\BibitemShut {NoStop}%
\bibitem [{\citenamefont {Dolde}\ \emph {et~al.}(2014)\citenamefont {Dolde},
  \citenamefont {Berholm}, \citenamefont {Wang}, \citenamefont {Jakobi},
  \citenamefont {Naydenov}, \citenamefont {Pezzagna}, \citenamefont {Meijer},
  \citenamefont {Jelezko}, \citenamefont {Neumann}, \citenamefont
  {Schulte-Herbrüggen}, \citenamefont {Biamonte},\ and\ \citenamefont
  {Wrachtrup}}]{Dolde2014}%
  \BibitemOpen
  \bibfield  {author} {\bibinfo {author} {\bibfnamefont {F.}~\bibnamefont
  {Dolde}}, \bibinfo {author} {\bibfnamefont {V.}~\bibnamefont {Berholm}},
  \bibinfo {author} {\bibfnamefont {Y.}~\bibnamefont {Wang}}, \bibinfo {author}
  {\bibfnamefont {I.}~\bibnamefont {Jakobi}}, \bibinfo {author} {\bibfnamefont
  {B.}~\bibnamefont {Naydenov}}, \bibinfo {author} {\bibfnamefont
  {S.}~\bibnamefont {Pezzagna}}, \bibinfo {author} {\bibfnamefont
  {J.}~\bibnamefont {Meijer}}, \bibinfo {author} {\bibfnamefont
  {F.}~\bibnamefont {Jelezko}}, \bibinfo {author} {\bibfnamefont
  {P.}~\bibnamefont {Neumann}}, \bibinfo {author} {\bibfnamefont
  {T.}~\bibnamefont {Schulte-Herbrüggen}}, \bibinfo {author} {\bibfnamefont
  {J.}~\bibnamefont {Biamonte}}, \ and\ \bibinfo {author} {\bibfnamefont
  {J.}~\bibnamefont {Wrachtrup}},\ }\href@noop {} {\bibfield  {journal}
  {\bibinfo  {journal} {Nat. Commun.}\ }\textbf {\bibinfo {volume} {5}},\
  \bibinfo {pages} {3371} (\bibinfo {year} {2014})}\BibitemShut {NoStop}%
\bibitem [{\citenamefont {Xia}\ \emph {et~al.}(2015)\citenamefont {Xia},
  \citenamefont {Lichtman}, \citenamefont {Maller}, \citenamefont {Carr},
  \citenamefont {Piotrowicz}, \citenamefont {Isenhower},\ and\ \citenamefont
  {Saffman}}]{Xia2015}%
  \BibitemOpen
  \bibfield  {author} {\bibinfo {author} {\bibfnamefont {T.}~\bibnamefont
  {Xia}}, \bibinfo {author} {\bibfnamefont {M.}~\bibnamefont {Lichtman}},
  \bibinfo {author} {\bibfnamefont {K.}~\bibnamefont {Maller}}, \bibinfo
  {author} {\bibfnamefont {A.}~\bibnamefont {Carr}}, \bibinfo {author}
  {\bibfnamefont {M.}~\bibnamefont {Piotrowicz}}, \bibinfo {author}
  {\bibfnamefont {L.}~\bibnamefont {Isenhower}}, \ and\ \bibinfo {author}
  {\bibfnamefont {M.}~\bibnamefont {Saffman}},\ }\href@noop {} {\bibfield
  {journal} {\bibinfo  {journal} {Phys. Rev. Lett.}\ }\textbf {\bibinfo
  {volume} {114}},\ \bibinfo {pages} {100503} (\bibinfo {year}
  {2015})}\BibitemShut {NoStop}%
\bibitem [{\citenamefont {Epstein}\ \emph {et~al.}(2014)\citenamefont
  {Epstein}, \citenamefont {Cross}, \citenamefont {Magesan},\ and\
  \citenamefont {Gambetta}}]{Epstein2014}%
  \BibitemOpen
  \bibfield  {author} {\bibinfo {author} {\bibfnamefont {J.}~\bibnamefont
  {Epstein}}, \bibinfo {author} {\bibfnamefont {A.}~\bibnamefont {Cross}},
  \bibinfo {author} {\bibfnamefont {E.}~\bibnamefont {Magesan}}, \ and\
  \bibinfo {author} {\bibfnamefont {J.}~\bibnamefont {Gambetta}},\ }\href@noop
  {} {\bibfield  {journal} {\bibinfo  {journal} {Phys. Rev. A}\ }\textbf
  {\bibinfo {volume} {89}},\ \bibinfo {pages} {062321} (\bibinfo {year}
  {2014})}\BibitemShut {NoStop}%
\bibitem [{\citenamefont {Gottesman}(1999)}]{Gottesman1999}%
  \BibitemOpen
  \bibfield  {author} {\bibinfo {author} {\bibfnamefont {D.}~\bibnamefont
  {Gottesman}},\ }in\ \href@noop {} {\emph {\bibinfo {booktitle} {Group22:
  Proceedings of the XXII International Colloquium on Group Theoretical Methods
  in Physics}}},\ \bibinfo {editor} {edited by\ \bibinfo {editor}
  {\bibfnamefont {S.}~\bibnamefont {Corney}}, \bibinfo {editor} {\bibfnamefont
  {R.}~\bibnamefont {Delbourgo}}, \ and\ \bibinfo {editor} {\bibfnamefont
  {P.}~\bibnamefont {Jarvis}}}\ (\bibinfo  {publisher} {Cambrige, MA,
  International Press},\ \bibinfo {year} {1999})\ pp.\ \bibinfo {pages}
  {32--43},\ \bibinfo {note} {arXiv:quant-ph/9807006}\BibitemShut {NoStop}%
\bibitem [{\citenamefont {Dankert}\ \emph {et~al.}(2009)\citenamefont
  {Dankert}, \citenamefont {Cleve}, \citenamefont {Emerson},\ and\
  \citenamefont {Livine}}]{Dankert2009}%
  \BibitemOpen
  \bibfield  {author} {\bibinfo {author} {\bibfnamefont {C.}~\bibnamefont
  {Dankert}}, \bibinfo {author} {\bibfnamefont {R.}~\bibnamefont {Cleve}},
  \bibinfo {author} {\bibfnamefont {J.}~\bibnamefont {Emerson}}, \ and\
  \bibinfo {author} {\bibfnamefont {E.}~\bibnamefont {Livine}},\ }\href@noop {}
  {\bibfield  {journal} {\bibinfo  {journal} {Phys. Rev. A}\ }\textbf {\bibinfo
  {volume} {80}},\ \bibinfo {pages} {012304} (\bibinfo {year}
  {2009})}\BibitemShut {NoStop}%
\bibitem [{\citenamefont {Mariantoni}\ \emph {et~al.}(2011)\citenamefont
  {Mariantoni}, \citenamefont {Wang}, \citenamefont {Yamamoto}, \citenamefont
  {Neeley}, \citenamefont {Bialczak}, \citenamefont {Chen}, \citenamefont
  {Lenander}, \citenamefont {Lucero}, \citenamefont {O'Connell}, \citenamefont
  {Sank}, \citenamefont {Weides}, \citenamefont {Wenner}, \citenamefont {Yin},
  \citenamefont {Zhao}, \citenamefont {Korotkov}, \citenamefont {Cleland},\
  and\ \citenamefont {Martinis}}]{Mariantoni2011}%
  \BibitemOpen
  \bibfield  {author} {\bibinfo {author} {\bibfnamefont {M.}~\bibnamefont
  {Mariantoni}}, \bibinfo {author} {\bibfnamefont {H.}~\bibnamefont {Wang}},
  \bibinfo {author} {\bibfnamefont {T.}~\bibnamefont {Yamamoto}}, \bibinfo
  {author} {\bibfnamefont {M.}~\bibnamefont {Neeley}}, \bibinfo {author}
  {\bibfnamefont {R.}~\bibnamefont {Bialczak}}, \bibinfo {author}
  {\bibfnamefont {Y.}~\bibnamefont {Chen}}, \bibinfo {author} {\bibfnamefont
  {M.}~\bibnamefont {Lenander}}, \bibinfo {author} {\bibfnamefont
  {E.}~\bibnamefont {Lucero}}, \bibinfo {author} {\bibfnamefont
  {A.}~\bibnamefont {O'Connell}}, \bibinfo {author} {\bibfnamefont
  {D.}~\bibnamefont {Sank}}, \bibinfo {author} {\bibfnamefont {M.}~\bibnamefont
  {Weides}}, \bibinfo {author} {\bibfnamefont {J.}~\bibnamefont {Wenner}},
  \bibinfo {author} {\bibfnamefont {Y.}~\bibnamefont {Yin}}, \bibinfo {author}
  {\bibfnamefont {J.}~\bibnamefont {Zhao}}, \bibinfo {author} {\bibfnamefont
  {A.}~\bibnamefont {Korotkov}}, \bibinfo {author} {\bibfnamefont
  {A.}~\bibnamefont {Cleland}}, \ and\ \bibinfo {author} {\bibfnamefont
  {J.}~\bibnamefont {Martinis}},\ }\href@noop {} {\bibfield  {journal}
  {\bibinfo  {journal} {Science}\ }\textbf {\bibinfo {volume} {334}},\ \bibinfo
  {pages} {61} (\bibinfo {year} {2011})}\BibitemShut {NoStop}%
\bibitem [{\citenamefont {Rebentrost}\ and\ \citenamefont
  {Wilhelm}(2009)}]{Rebentrost2009}%
  \BibitemOpen
  \bibfield  {author} {\bibinfo {author} {\bibfnamefont {P.}~\bibnamefont
  {Rebentrost}}\ and\ \bibinfo {author} {\bibfnamefont {F.}~\bibnamefont
  {Wilhelm}},\ }\href@noop {} {\bibfield  {journal} {\bibinfo  {journal} {Phys.
  Rev. B}\ }\textbf {\bibinfo {volume} {79}},\ \bibinfo {pages} {060507(R)}
  (\bibinfo {year} {2009})}\BibitemShut {NoStop}%
\bibitem [{\citenamefont {Perron}(1907)}]{Perron1907}%
  \BibitemOpen
  \bibfield  {author} {\bibinfo {author} {\bibfnamefont {O.}~\bibnamefont
  {Perron}},\ }\href@noop {} {\bibfield  {journal} {\bibinfo  {journal} {Math.
  Ann.}\ }\textbf {\bibinfo {volume} {64}},\ \bibinfo {pages} {248} (\bibinfo
  {year} {1907})}\BibitemShut {NoStop}%
\bibitem [{\citenamefont {Frobenius}(1912)}]{Frobenius1912}%
  \BibitemOpen
  \bibfield  {author} {\bibinfo {author} {\bibfnamefont {G.}~\bibnamefont
  {Frobenius}},\ }\href@noop {} {\bibfield  {journal} {\bibinfo  {journal}
  {Sitzungsber. K\"onigl. Preuss. Akad. Wiss.}\ }\textbf {\bibinfo {volume}
  {-}},\ \bibinfo {pages} {456} (\bibinfo {year} {1912})}\BibitemShut {NoStop}%
\bibitem [{\citenamefont {Meyer}(2000)}]{Meyer2000}%
  \BibitemOpen
  \bibfield  {author} {\bibinfo {author} {\bibfnamefont {C.}~\bibnamefont
  {Meyer}},\ }\href@noop {} {\emph {\bibinfo {title} {Matrix analysis and
  applied linear algebra}}}\ (\bibinfo  {publisher} {SIAM},\ \bibinfo {year}
  {2000})\ p.\ \bibinfo {pages} {655}\BibitemShut {NoStop}%
\bibitem [{\citenamefont {Wallman}\ \emph {et~al.}(2014)\citenamefont
  {Wallman}, \citenamefont {Barnhill},\ and\ \citenamefont
  {Emerson}}]{Wallman2014X}%
  \BibitemOpen
  \bibfield  {author} {\bibinfo {author} {\bibfnamefont {J.}~\bibnamefont
  {Wallman}}, \bibinfo {author} {\bibfnamefont {M.}~\bibnamefont {Barnhill}}, \
  and\ \bibinfo {author} {\bibfnamefont {J.}~\bibnamefont {Emerson}},\
  }\href@noop {} {\enquote {\bibinfo {title} {Characterization of leakage
  errors via randomized benchmarking},}\ } (\bibinfo {year} {2014}),\ \bibinfo
  {note} {arXiv:1412.4126}\BibitemShut {NoStop}%
\bibitem [{\citenamefont {C\'orcoles}\ \emph {et~al.}(2013)\citenamefont
  {C\'orcoles}, \citenamefont {Gambetta}, \citenamefont {Chow}, \citenamefont
  {Smolin}, \citenamefont {Ware}, \citenamefont {Strand}, \citenamefont
  {Plourde},\ and\ \citenamefont {Steffen}}]{Corcoles2013}%
  \BibitemOpen
  \bibfield  {author} {\bibinfo {author} {\bibfnamefont {A.}~\bibnamefont
  {C\'orcoles}}, \bibinfo {author} {\bibfnamefont {J.}~\bibnamefont
  {Gambetta}}, \bibinfo {author} {\bibfnamefont {J.}~\bibnamefont {Chow}},
  \bibinfo {author} {\bibfnamefont {J.}~\bibnamefont {Smolin}}, \bibinfo
  {author} {\bibfnamefont {M.}~\bibnamefont {Ware}}, \bibinfo {author}
  {\bibfnamefont {J.}~\bibnamefont {Strand}}, \bibinfo {author} {\bibfnamefont
  {B.}~\bibnamefont {Plourde}}, \ and\ \bibinfo {author} {\bibfnamefont
  {M.}~\bibnamefont {Steffen}},\ }\href@noop {} {\bibfield  {journal} {\bibinfo
   {journal} {Phys. Rev. A}\ }\textbf {\bibinfo {volume} {87}},\ \bibinfo
  {pages} {030301} (\bibinfo {year} {2013})}\BibitemShut {NoStop}%
\bibitem [{\citenamefont {Wallman}\ and\ \citenamefont
  {Flammia}(2014)}]{Wallman2014}%
  \BibitemOpen
  \bibfield  {author} {\bibinfo {author} {\bibfnamefont {J.}~\bibnamefont
  {Wallman}}\ and\ \bibinfo {author} {\bibfnamefont {S.}~\bibnamefont
  {Flammia}},\ }\href@noop {} {\bibfield  {journal} {\bibinfo  {journal} {New.
  J. Phys.}\ }\textbf {\bibinfo {volume} {16}},\ \bibinfo {pages} {103032}
  (\bibinfo {year} {2014})}\BibitemShut {NoStop}%
\end{thebibliography}%
\bibliographystyle{apsrev4-1}
\end{document}